\documentclass[%
prd, reprint,
nofootinbib,
amsmath,amssymb,
aps, prd,
floatfix,
]{revtex4-1}

\usepackage{subcaption}
\usepackage{tensor}
\usepackage{graphicx}
\usepackage{dcolumn}
\usepackage{bm}
\usepackage{hyperref}
\usepackage{natbib}
\usepackage{color}

\usepackage{chngcntr}
\counterwithin{paragraph}{subsection}

\begin{document}

\title{Black hole energy extraction via stationary scalar clouds}%

\author{Jordan Wilson-Gerow$^{1,2}$ and Adam Ritz$^{1}$}
\affiliation{$^{1}$Department of Physics \& Astronomy, University of Victoria,
Victoria, BC V8W 3P6, Canada\\
$^{2}$Department of Physics \& Astronomy, University of British Columbia, 
Vancouver, BC V6T 1Z1, Canada}

\date{September 2015}

\begin{abstract}
\noindent
We study scalar field configurations around Kerr black holes with a time-independent energy-momentum tensor. These stationary `scalar clouds', confined near the black hole (BH) by their own mass or a mirror at fixed radius, exist at the threshold for energy extraction via superradiance. Motivated by the electromagnetic Blandford-Znajek (BZ) mechanism, we explore whether scalar clouds could serve as a proxy for the force-free magnetosphere in the BZ process. 
We find that a stationary energy-extracting scalar cloud solution exists when the reflecting mirror is replaced by a semi-permeable surface which allows the cloud to radiate some energy to infinity while maintaining self-sustained superradiance. The radial energy flux displays the same behaviour for rapidly rotating holes as magnetohydrodynamic simulations predict for the BZ mechanism.
\end{abstract}

\maketitle

\section{\label{sec:Introduction}Introduction}
 Black hole energy extraction has been a topic of great interest since the discovery of active galactic nuclei (AGN). These ultra-compact regions at the center of many galaxies are the most energetic phenomena in the universe. It is generally accepted that AGN are driven by supermassive black holes at the center of these galaxies, but the details of the power source are still debated. Studies of black hole energy extraction date back to Penrose's \textit{gedanken}  experiment in which a clump of matter disintegrates within the ergoregion of a Kerr black hole~\cite{Penrose:1971uk}. While the Penrose process demonstrates the possibility of classical energy extraction from rotating black holes, it is unlikely to be significant on astrophysical scales. 
 Nonetheless, dynamics within the ergoregion remains the essential component of all classical processes of this type. 
 
Superradiance is another mechanism for energy extraction in which bosonic waves of the form $e^{i(m\phi-\omega t)}$ are scattered off a black hole rotating with angular speed $\Omega_{H}=\frac{a}{r_{+}^2+a^2}$. If the waves satisfy the superradiant condition $\omega<m\Omega_{H}$ then they are reflected and amplified rather than absorbed~\cite{Bekenstein:1973mi, PhysRevLett.28.994, JETP.Let.14.180}. This can be interpreted as a consequence of the second law of black hole thermodynamics. Classically, the area of a black hole must increase~\cite{Hawking:1971vc}.  When modes satisfying the superradiant condition impinge on a BH, the hole must give up some energy to ensure that the area increases~\cite{Bekenstein:1973mi, Brito:2015oca},
\begin{equation} \label{eq:areaRel}
\delta M=\frac{\omega T_{H}}{4}\frac{\delta A_{H}}{\omega-m\Omega_{H}}.
\end{equation}
A more physically intuitive picture of rotational superradiance is to consider two zero angular momentum observers (ZAMOs), one very far from the hole and one very near the horizon. From the viewpoint of the observer very far from the hole, only positive (prograde) angular momentum is dumped onto the black hole. Due to frame dragging, the near-ZAMO rotates with the horizon angular speed $\Omega_{H}$. The relative angular velocity of the near horizon ZAMO and the incident waves is $\omega/m-\Omega_{H}$, so the near-horizon ZAMO may observe the black hole absorbing negative angular momentum waves.  Due to the existence of an ergoregion in stationary axisymmetric spacetimes, the near-horizon observer and the observer at infinity disagree about what is considered energy and what is angular momentum.  For superradiant modes, while the near-horizon ZAMO sees positive energy and negative angular momentum impinging on the hole, the distant observer instead sees positive angular momentum and negative energy impinging on the hole. Thus, energy as measured at infinity is extracted from the hole by the scattering of superradiant modes.

To construct a configuration in which energy is continuously extracted via superradiant scattering (as one would desire for a model of AGN) requires an influx of waves from infinity.  This boundary condition does not correspond to a self-sustaining model of energy extraction. Instead, the most astrophysically plausible mechanism was proposed by Blandford and Znajek ~\cite{Blandford:1977ds}.  The BZ mechanism involves a force-free magnetosphere surrounding a Kerr black hole.  The frame-dragging associated with the black hole perturbs the electromagnetic field configuration and generates an outward-directed Poynting flux. While the BZ mechanism is a prime candidate to generate AGN power, the field equations for a magnetically dominant force-free magnetosphere have only been solved analytically in the perturbative slow-rotation limit~\cite{Blandford:1977ds, Tanabe:2008wm}. Much of the research into the BZ mechanism involves general relativistic magnetohydrodynamic (GRMHD) simulations~\cite{Komissarov:2004qu, McKinney:2004ka, Tchekhovskoy:2009ba}.  Historically, GRMHD simulations have disagreed with the perturbative results in the limit of rapid rotation, motivating further attempts to understand energy extraction in the near-extremal regime analytically. Indeed, recent work has claimed to extend the perturbative expansion to higher order~\cite{Pan:2015iaa}, leading to consistency with numerical results over the entire range of black hole spin.

Although rather different in appearance, the BZ mechanism and superradiance are fundamentally quite similar. Both mechanisms necessarily rely on the existence of an ergoregion outside the black hole. The condition for energy extraction via superradiance is $\frac{\omega}{m}<\Omega_{H}$, where $\frac{\omega}{m}$ is the angular speed of the scalar field. Similarly, the condition for energy extraction via the BZ mechanism is $\Omega_{EM}<\Omega_{H}$, where $\Omega_{EM}$ is the angular speed of the electromagnetic field.  Furthermore for a scalar field $\Phi$ and a force-free electromagnetic gauge field $A_{\mu}$, the expressions for the energy flux through the horizon are respectively $-2\omega(\omega-m\Omega_{H})|\Phi|^{2}$ and $-\Omega_{EM}(\Omega_{EM}-\Omega_{H})(\partial A_{\phi})^{2}$. As a consequence, the above discussion of superradiance also applies to the BZ mechanism if one replaces the quantities $(\omega,m)$ with $(\Omega_{EM},1)$. Nonetheless, despite this fundamental similarity, the quantitative details of the BZ mechanism are obscured by the complexity of the fundamental equations describing the BZ mechanism.\footnote{Recently, a class of exact solutions for force-free magnetospheres has been constructed in the special limit of a null current \cite{BGJ,MD}.}   As a result, the quantitative details of scalar superradiance are more accessible, with an analytic solution to the field equation available for the entire range of black hole spins, $0<a<M$. However, as noted above, an important distinction with the BZ mechanism is that superradiant solutions do not lead to a stationary energy flux from the BH. Our goal in this paper is to utilize scalar superradiance as a tractable proxy for the BZ mechanism, by obtaining a physical setting in which the energy flux is time-independent. In particular, a solution available over the entire range of black hole spin would be valuable in allowing a comparison with GRMHD results for the BZ mechanism, particularly for near-extremal solutions.

The rest of this paper is organized as follows.  In Section~2 we discuss solutions for scalar fields on Kerr backgrounds dubbed \textit{scalar clouds}~\cite{Hod:2012px,Herdeiro:2014goa}, that fluctuate with a frequency right at the threshold for superradiance, and are `stationary' in the specific sense of having a time-independent energy-momentum tensor. We proceed to study the effect of perturbations of these scalar clouds with the intention of provoking the cloud to support time-independent energy extraction from the BH. In Section~3 we present a analytic solution that utilizes a semi-permeable membrane outside the ergosphere, and compare the energy extraction rates to those obtained from the BZ mechanism. We conclude in Section~4 with some remarks on backreaction and further applications.

\section{\label{sec:Scalar Cloud}Scalar Clouds}

Working within the probe approximation, we neglect the backreaction of the scalar field on the metric. The vacuum spacetime of interest is the Kerr black hole. In Boyer-Lindquist coordinates $\{t,r,\theta,\phi\}$, the line element takes the form
\begin{align}
ds^2 &= - \left(1-\frac{2r}{\Sigma}\right)dt^2 + \frac{\Sigma}{\Delta} dr^2 + \Sigma d\theta^2 \nonumber\\
& \qquad + \frac{A \sin^2\theta}{\Sigma} d\phi^2 - \frac{4ar \sin^2\theta}{\Sigma} d\phi dt,
\end{align}
where $\Sigma \equiv r^2 + a^2\cos^2\theta$, $\Delta \equiv r^2 - 2r+a^2$ and $A\equiv (r^2+a^2)^2- a^2 \Delta \sin^2\theta$. In these coordinates, the metric is singular on the horizon at $r=r_+$ where $\Delta =0$. When required, we will also make use of horizon penetrating Kerr-Schild coordinates. Throughout, we work with Planck units, where $G=c=1$.

The action for a complex scalar field $\Phi$ with mass $\mu$ takes the form
\begin{equation}
S=\int d^4 x\, \sqrt[]{-g} \left( -\nabla_{\mu}\Phi^{*} \nabla^{\mu}\Phi-\mu^{2}\Phi^{*}\Phi \right),
\end{equation}
leading to the Klein-Gordon equation of motion,
\begin{equation}
\left(\nabla^{\mu}\nabla_{\mu}-\mu^{2}\right)\Phi=0.
\end{equation}

Exploiting the timelike and axial symmetries of the background, we consider oscillatory solutions of the form,
\begin{equation}
 \Phi=\chi(r,\theta)e^{im\phi}e^{-i\omega t}.
 \end{equation}
When the scalar field is sufficiently massive $(\mu^{2}>\omega^{2})$, and the frequency is equal to the critical superradiant frequency, 
\begin{equation}
 \omega = \omega_c \equiv m\Omega_{H}, 
 \end{equation}
 a stationary bound state of the scalar field, or scalar cloud, exists~\cite{Hod:2012px}. These solutions were first obtained in the probe limit around extremal black holes, but have since been shown to persist as full solutions of the Einstein-scalar system. They bypass the no-hair theorem through specific time-dependence,  that still leads to a stationary energy-momentum tensor \cite{Herdeiro:2014goa}. In accord with the previous physical picture, the critical frequency is analogous to the rotation speed at which the near-horizon ZAMO sees no angular momentum impinging on the BH.  Scalar clouds owe their existence to a resonance effect in which they neither extract energy from nor lose energy to the black hole. While they do not extract any energy from the black hole, they provide a medium which persists in the ergosphere, similar to the physical magnetosphere. In the BZ mechanism, a force-free magnetosphere supports time-independent energy extraction from the black hole. The scalar system exhibits a cloud solution that does not support energy extraction on its own, but the system is on the verge of superradiance.  In the next section, we explore whether the cloud can be perturbed to support time-independent energy extraction, analogous to the force-free magnetosphere in the BZ mechanism.

\subsection{\label{sec:Perturbations}Perturbations}

The complex scalar system can be perturbed in numerous ways by adding various source or interaction terms to the action.  However as long as the source terms respect the symmetry of the background spacetime (stationarity and axisymmetry), the equation of motion will always possess a solution of the form $\Phi=\chi(r,\theta)e^{im\phi}e^{-i\omega t}$. 
Due to the $U(1)$ symmetry of the theory, this statement also holds in the presence of interactions, e.g.~$|\Phi|^4$, which respect that symmetry.  

To investigate the effect of these perturbations on energy extraction from the black hole we must consider the conserved energy flux for the scalar field, $j^{\mu}=-\tensor{T}{^{\mu}_{\nu}}k^{\nu}$, where $T^{\mu\nu}$ is the energy momentum tensor of the scalar field and $k^{\nu}$ is the Killing vector which is timelike at infinity. To calculate the rate of energy extraction from the black hole we project the energy flux vector along the null horizon generating Killing vector $\xi^{\mu}_{\mathcal{H}}=k^{\mu}+m^{\mu}\Omega_{H}$, where $m^{\mu}$ is the azimuthal Killing vector.  The outgoing energy flux through the horizon is given by the expression 
\begin{equation} 
\dot{E}=\int_{\mathcal{H}\cap\Sigma_{t}}dS\, \xi^{\mathcal{H}}_{\mu}j^{\mu},
\end{equation}
where $\Sigma_{t}$ is a constant time slice.

For the scalar ansatz above, this expression becomes
\begin{equation}
\dot{E}=-2\int_{\mathcal{H}\cap\Sigma_{t}} dS\, |\Phi|^{2} \left[ \Im(\omega)^2 +\Re(\omega) (\Re(\omega)-m\Omega_{H}) \right].
\end{equation}
The addition of stationary axisymmetric sources can only modify the weight $|\Phi|^{2}$, and thus no sources respecting the background symmetry can perturb a scalar cloud solution to support energy extraction.

If instead, we minimally couple the complex scalar field to a background  gauge field $A_{\mu}$, it is $p_{t}=\omega+qA_{t}$ and $p_{\phi}=m-qA_{\phi}$ which enter the expression for the horizon flux. For a scalar field with charge $q$ coupled to a gauge field $A_{\mu}$ the outgoing energy flux through the horizon is 
\begin{equation} \label{eq:coupledFlux}
\dot{E}=-2\int_{\mathcal{H}\cap\Sigma_{t}} dS\, |\Phi|^{2} \left[ \Im(p_{t})^2{+}(\Re(p_{t})) (\Re(p_{t}){-}p_{\phi}\Omega_{H}) \right]. 
\end{equation}

It follows that a scalar cloud characterized by $\omega=m\Omega_{H}$ can be  perturbed into the energy extracting regime by coupling the scalar field to a horizon penetrating background gauge field. Such charged solutions have been discussed in \cite{HodEM,BenoneEM}. However, such a coupling effectively redefines the scalar frequency and angular momentum, which moves the superradiant critical frequency in parameter space. Thus, introducing this coupling has the effect of pushing the scalar field solution away from the stationary bound state, and back into the super radiant regime. Thus, while energy is extracted, it does not meet the goal of achieving a configuration in which the flux is time independent.

The stationarity of the bound scalar cloud was of specific interest as a potential proxy for the stationary electromagnetic field in the BZ mechanism. Instead, we have found that local perturbations tend only to alter the tomography of the cloud and/or redefine the critical cloud frequency.  In the next section, we explore another approach focusing more directly on the superradiant regime.

\section{\label{sec:Confinement} Energy Extraction via Partial Confinement}

As is well known from studies of superradiance, scalar fields with frequencies not equal to the critical frequency can potentially grow exponentially~\cite{Press:1972zz}. This phenomena, proposed by Teukolsky and Press, is called the \textit{black hole bomb} and it occurs when superradiant radiation is continually reflected back onto the black hole. In this section, we reconsider this mechanism and obtain a stationary variant via a global change to the boundary conditions on the reflecting surface.

\subsection{Confinement and the Superradiant Instability}

 The onset of the instability requires two factors, superradiance and confinement~\cite{Press:1972zz}. There are many known methods to confine scalar radiation, \textit{e.g.} mirror-like reflecting boundary conditions~\cite{Cardoso:2004nk}, an anti-de Sitter geometry~\cite{PhysRevD.70.084011}, strong magnetic fields~\cite{PhysRevD.89.104045}, and ``mass mirrors"~\cite{damour, Hod:2009cp}. The bound nature of the scalar cloud follows when its mass satisfies the mass mirror condition $\mu^{2}>\omega^{2}$. Asymptotically, the scalar field decays exponentially as $e^{-\,\sqrt[]{\mu^{2}-\omega^{2}}r}$, so any energy extracted from the black hole in the form of scalar radiation cannot radiate to infinity and the amplitude of the scalar field must grow.  If we were to allow the radiation to escape to infinity by considering modes with sufficiently large frequency $\omega^{2}>\mu^{2}$, then the scalar field could no longer form a bound state and the energy extraction mechanism would not be self-sustaining.

The instability can be characterized by focusing on the conservation of energy. Take a region of spacetime $\mathcal{S}$ bounded by two constant time hypersurfaces $\Sigma_{t_{1}}$ and $\Sigma_{t_{2}}$ $(t_{2}=t_{1}+dt)$, and two constant radius hypersurfaces $\Sigma_{\mathcal{H}}$ and $\Sigma_{\infty}$ (the horizon and an asymptotic surface respectively). Applying Gauss' theorem we obtain the relation
\begin{align} \label{eq:consEng}
0 &=\int_{\mathcal{S}}d^{4}x \,\sqrt[]{-g}\,\nabla_{\mu}j^{\mu}      \nonumber\\
 &=\left(\int_{\Sigma_{t_{2}}}{-}\int_{\Sigma_{t_{1}}} \right) d^3 x\,\sqrt[]{^{3}g}\, k_{\mu}j^{\mu}\nonumber\\
  & \qquad \qquad \qquad - \int_{\Sigma_{\mathcal{H}}}dS_{\mu}j^{\mu}{+}\int_{\Sigma_{\infty}}dS_{\mu}j^{\mu},
\end{align}
where $dS_{\mu}$ is the restricted volume element for the boundary $\partial \mathcal{S}$. 

The first term in (\ref{eq:consEng}) is simply the total energy change between the spacelike hypersurfaces. Since the constant time slices are infinitesimally separated we can rewrite (\ref{eq:consEng}) as
\begin{equation}
0=\frac{dE}{dt}-\int_{\mathcal{H}\cap\Sigma_{t}}dA \xi^{\mathcal{H}}_{\mu}j^{\mu}+Q, \label{eq:cons2}
\end{equation}
where Q is the rate of energy lost to infinity.

 The remaining integral in (\ref{eq:cons2}) is the outgoing energy flux through the horizon, \textit{i.e.} the rate of energy extraction from the black hole.  For a free scalar field with time dependence $e^{-i\omega t}$ the energy momentum tensor will be proportional to $e^{2\Im(\omega) t }$, so that the change in total energy is given by $\frac{dE}{dt}=2\Im(\omega)E$. This quantifies the intuitive notion that if no energy is lost to infinity but energy flows across the horizon, then the scalar frequency is necessarily complex and the scalar field is unstable to either collapse or the black hole bomb process. Clearly the only way to avoid the instability is to allow the field to dissipate some energy to infinity. It follows that a stable self-sustaining mechanism may follow by introducing partial reflection off a semipermeable surface outside the hole. If an energy extracting scalar field decays exponentially in the absence of a confining mechanism but grows exponentially in the presence of full confinement, then continuity requires there to exist some degree of partial confinement for which the field will be stationary but also extract energy from the black hole.

\subsection{\label{sec:Partial}Partial Confinement}

The aforementioned system can be analyzed in a manner very similar to the study of confinement via a perfectly reflecting mirror~\cite{Cardoso:2004nk}. To proceed analytically we assume that the semipermeable surface is spherical and located far from the hole at $r_{m}\gg M$. For simplicity, we will assume a infinitely thin non-dynamical surface, localized as $\delta(r-r_{m})$. We will also consider the case in which the scalar field is both massless and electrically neutral. We could consider a massive field and the results of the following discussion would be qualitatively similar provided that $\omega^{2}>\mu^{2}$.

The equation of motion is now
\begin{equation}
\nabla^{\mu}\nabla_{\mu}\Phi=0,
\end{equation}
which in the Kerr background is separated with the ansatz $\Phi=R(r)S(\theta)e^{im\varphi}e^{-i\omega t}$ into a radial ODE and an angular ODE coupled by a constant,
\begin{equation}
\frac{1}{\sin\theta}\frac{d}{d\theta}\left(\sin\theta \frac{dS}{d\theta}\right)+\left(K+a^2 \omega^2 \cos\theta -\frac{m^2}{\sin^{2}\theta} \right)S=0,
\end{equation}
and
\begin{equation}
\Delta \frac{d}{dr}\left( \Delta \frac{dR}{dr} \right){+}((\omega(r^2{+}a^2){-}ma)^{2}{+}\Delta (2ma\omega{-}K))R{=}0.
\end{equation}
The angular equation is solved by the spheriodal harmonics $S_{lm}$, and the coupling constant $K_{lm}=l(l+1)+\mathcal{O}(a^2 \omega^2)$ is given in~\cite{abramowitz}.

At the horizon, semipermeable surface, and spatial infinity we must impose boundary conditions on the radial solution. On the horizon, the relevant boundary condition is to impose that there are purely ingoing waves.  Just to the inside of the semipermeable surface we allow a linear combination of ingoing and outgoing waves, and approaching spatial infinity there should be purely outgoing waves,
\begin{equation}
    \Phi\sim
\begin{cases}
    \frac{1}{r} e^{-i\omega r_{*}}e^{-i\omega t},&  r\to \infty,\\
    e^{-i(\omega-m\Omega_{H})r_{*}}e^{-i\omega t}, &r\to r_{+},
\end{cases}
\end{equation}
where $r_{+}$ is the tortise coordinate defined by $\frac{dr_{*}}{dr}=\frac{r^2+a^2}{\Delta}$. Note that on the horizon, while the group velocity is negative, the phase velocity is positive for frequencies in the superradiant regime. 

In the low energy limit, $\omega M\ll1$, the radial equation is amenable to the method of matched asymptotics. In the near-field region, $(r-r_{+})\leq\frac{1}{\omega}$, the radial equation reduces to
\begin{equation}
 \Delta \frac{d}{dr}\left( \Delta \frac{dR}{dr} \right){+}\left( (r_{+}^2{+}a^2)^{2}(\omega{-}m\Omega_{H})^{2}{-}\Delta l(l{+}1) \right)R{=}0.
\end{equation}
The solution, corresponding to ingoing waves at the horizon, is a hypergeometric function, 
\begin{equation} \label{eq:nearSol}
R=Az^{-i\overline{\omega}}(1-z)^{l+1}F(a-c+1,b-c+1,2-c,z)
\end{equation}
where $z=\frac{r-r_{+}}{r-r_{-}}$ and, following \cite{Cardoso:2004nk}, we have defined the superradiant factor
\begin{equation}
\overline{\omega}=(\omega-m\Omega_{H})\frac{r_{+}^{2}+a^{2}}{r_{+}-r_{-}}.
\end{equation}
The arguments of $F$ include $a=l+1+2i\overline{\omega}$, $b=l+1$, and $c=1+2i\overline{\omega}$. 

In the far-field region, $r-r_{+}\gg M$, spacetime is approximately flat so the radial equation reduces to
\begin{equation}
\frac{d}{dr}\left( r^2 \frac{dR}{dr} \right)+\left( r^{2}\omega^{2} - l(l+1) \right)R=0.
\end{equation}
The general solution can be written as a linear combination of ingoing and outgoing Hankel functions,
\begin{equation} \label{eq:farSol}
R=r^{-\frac{1}{2}}\left( \alpha H^{(1)}_{l+\frac{1}{2}}(\omega r) +\beta H^{(2)}_{l+\frac{1}{2}}(\omega r) \right).
\end{equation}

In the intermediate region, $M\ll r-r_{+}\ll \frac{1}{\omega}$, one can match the large r expansion of (\ref{eq:nearSol}) to the small r expansion of (\ref{eq:farSol}).  The matching procedure \cite{Cardoso:2004nk} relates the coefficients of the solutions,
\begin{equation} \label{eq:normRelation}
A=\frac{(r_{+}-r_{-})^l \Gamma(l+1) \Gamma(l+1-2i\overline{\omega})}{\Gamma(l+\frac{3}{2}) \Gamma(2l+1)\Gamma(1-i2\overline{\omega})}\left( \frac{\omega}{2} \right)^{l+\frac{1}{2}}(\alpha +\beta)
\end{equation}
where
\begin{equation}
\frac{\beta}{\alpha}=\frac{1+(\omega-m\Omega_{H})\gamma}{1-(\omega-m\Omega_{H})\gamma},
\end{equation}
in terms of the parameter  
\begin{align}
\gamma &=\frac{2(r_{+}^{2}+a^{2})}{2l+1}\left(\frac{l!}{(2l-1)!!} \right)^2   \frac{(r_{+}-r_{-})^{2l}}{(2l)!(2l+1)!} \nonumber\\
& \qquad\qquad \times \left(\prod_{k=1}^l (k^2+4\overline{\omega}^2) \right) \omega^{2l+1}.
\end{align}
This fixes all the coefficients in terms of one parameter, e.g. $\alpha$,  the normalization of the outgoing solution in the far-field region.

To impose the semipermeable surface condition at $r=r_{m}$ we demand that outside the surface the solution is purely outgoing and also that $\Phi$ is continuous at $r=r_{m}$. Letting $\alpha\mathcal{T}$ be the amplitude of the transmitted wave, we arrive at the relationship
\begin{equation}
\frac{\beta}{\alpha}=(\mathcal{T}-1)\frac{H^{(1)}_{l+\frac{1}{2}}(\omega r_m)}{H^{(2)}_{l+\frac{1}{2}}(\omega r_m)}.
\end{equation}

In the far-field region, $\omega r_{m}\gg1$, the solutions are spherical waves with radial dependence $\propto \frac{1}{r} e^{\pm i\omega r}$. Thus the ratio of Hankel functions simplifies to an exponential function $(-1)^{l+1} e^{2ir_{m}\omega}$. Bringing together the matched solution constraint as well as the constraint of continuity across the semipermeable surface we arrive at an equation relating the transmission coefficient $\mathcal{T}$ to the frequency,
\begin{equation}
(\mathcal{T}-1)(-1)^{l+1} e^{2ir_{m}\omega}=\frac{1+(\omega-m\Omega_{H})\gamma}{1-(\omega-m\Omega_{H})\gamma}. \label{Tw}
\end{equation}
This is a transcendental equation for $\omega$, but in the present regime with $\omega M \ll 1$, the solution to first order in $\omega$ takes the following form when 
$\gamma\approx 0$,
\begin{equation}
\omega_0=\frac{\pi}{2r_m}(l+2n)+i \frac{\log(1-\mathcal{T})}{2r_m}
\end{equation}
for $n\in \{ 0,1,2,..\}$.

We now treat the problem perturbatively to first order in $\gamma$. Assuming a solution of the form $\omega=\omega_{0}+i\delta$ where $\delta\ll\omega_{0}$, we can Taylor expand (\ref{Tw}) to obtain
\begin{equation}
\delta=-(\Re(\omega_{0})-m\Omega_{H})\frac{\gamma}{r_{m}} + {\cal O}(\gamma^2).
\end{equation}
Thus the complete solution for $\omega$ in the limit $\omega M\ll1$ is approximately
\begin{align}
\omega &=\frac{\pi}{2r_m}(l+2n) \nonumber\\
 & \qquad +\frac{i}{2r_{m}} \left( \log(1-\mathcal{T})-2(\Re(\omega)-m\Omega_{H})\gamma \right).
\end{align}

Now that we have obtained the quasinormal modes of this system we can find the transmission coefficient necessary to ensure the stationarity of the scalar field by setting  $\Im (\omega)=0$,
\begin{equation}
\mathcal{T}=1-e^{2\gamma(\omega-m\Omega_{H})}.
\end{equation}
In the $\mathcal{T}\to 0$ limit (perfect reflection) we see that the only stationary solution is the original scalar cloud, $\omega=m\Omega_{H}$, as expected.

The energy momentum tensor for this configuration is completely time independent. The outward energy flux through the horizon is strongly dependent on the scalar frequency $\omega$ and horizon angular velocity $\Omega_{H}$,
\begin{equation}
\dot{E}=-4\pi(r_{+}^2+a^2 ) \omega(\omega-m\Omega_{H})|A|^{2}\int  d\theta \sin\theta |S(\theta)|^{2}.
\end{equation}

By the continuity and stationarity of the scalar field, all of the energy flux out of the horizon must be carried away by a conserved radial current, $i(\Phi\partial_{r}\Phi^{*}-\Phi^{*}\partial_{r}\Phi)$. As a consistency check, we calculate the radial energy flux on the horizon $\mathcal{F}^{r}_{E}\equiv \int_{\mathcal{H}\cap \Sigma_{t}} dS\, n_{\mu}j^{\mu}$ where  $n_{\mu} \sim \nabla_{\mu}r$ is the radial normal vector,
\begin{equation}
\mathcal{F}^{r}_{E}= -2\pi i\omega(r_{+}^{2}+a^{2})\Delta\int d\theta \sin\theta (\Phi\partial_{r}\Phi^{*}-\Phi^{*}\partial_{r}\Phi).
\end{equation}

Differentiating (\ref{eq:nearSol}) and evaluating the expression on the horizon we find the radial energy carried away from the horizon precisely matches the energy flux through the horizon,
\begin{equation}
\mathcal{F}^{r}_{E}=-4\pi(r_{+}^2+a^2 ) \omega(\omega-m\Omega_{H})|A|^{2}\int  d\theta \sin\theta |S(\theta)|^{2}.
\end{equation}

Thus the extracted energy is indeed carried away from the black hole in the form of a scalar current. We can proceed to ask how the rate of energy extraction depends on the rotation speed of the black hole. Using (\ref{eq:normRelation}), the near-region normalization constant can be expressed in terms of the far-region normalization constant. The full expression for the energy extraction rate then takes the form,
\begin{align}
\dot{E} &= -2\pi |\alpha|^2  (r_{+}^{2}+a^{2}) (\frac{1}{2})^{2l}\omega^{2l+2}(\omega-m\Omega_{H})  \nonumber\\
 & \quad \times  \prod\limits_{k=1}^{l}\left[k^{2}(r_{+}-r_{-})^{2}+4(\omega-m\Omega_{H})^{2}(r_{+}^{2}+a^{2})^{2} \right] \nonumber \\
 & \quad \times \frac{\Gamma(l+1)}{\Gamma(l+\frac{3}{2})\Gamma(2l+1)}\int d\theta \sin\theta |S(\theta)|^{2}. \label{Edot}
\end{align}

It is interesting to compare this fully analytic result with the numerical and perturbative results for the radial energy flux sourced by the BZ magnetosphere. Ideally, it would be useful to normalize the flux by the energy on the horizon for example, $\int dS\, T_{\mu \nu}k^{\mu}k^{\nu}$, to factor out the arbitrary overall amplitude. Unfortunately, in Kerr-Schild coordinates, the electromagnetic energy density vanishes on the horizon in the extremal limit. Thus, we will instead simply compare the qualitative shapes of the $\dot{E}(\Omega_{H})$ curves, by normalizing the fields such that the maximum energy extraction rates are equal. This provides no information about the relative efficiency, but it still allows one to compare the features of the curves. The BZ curve we use in this comparison is the ${\cal O}(\Omega_H^8)$ approximation to the split monopole magnetosphere obtained in~\cite{Pan:2015iaa}, which has been shown to closely match the results of numerical simulations.

\begin{figure}[t]
\centering
  \includegraphics[width=0.95\linewidth]{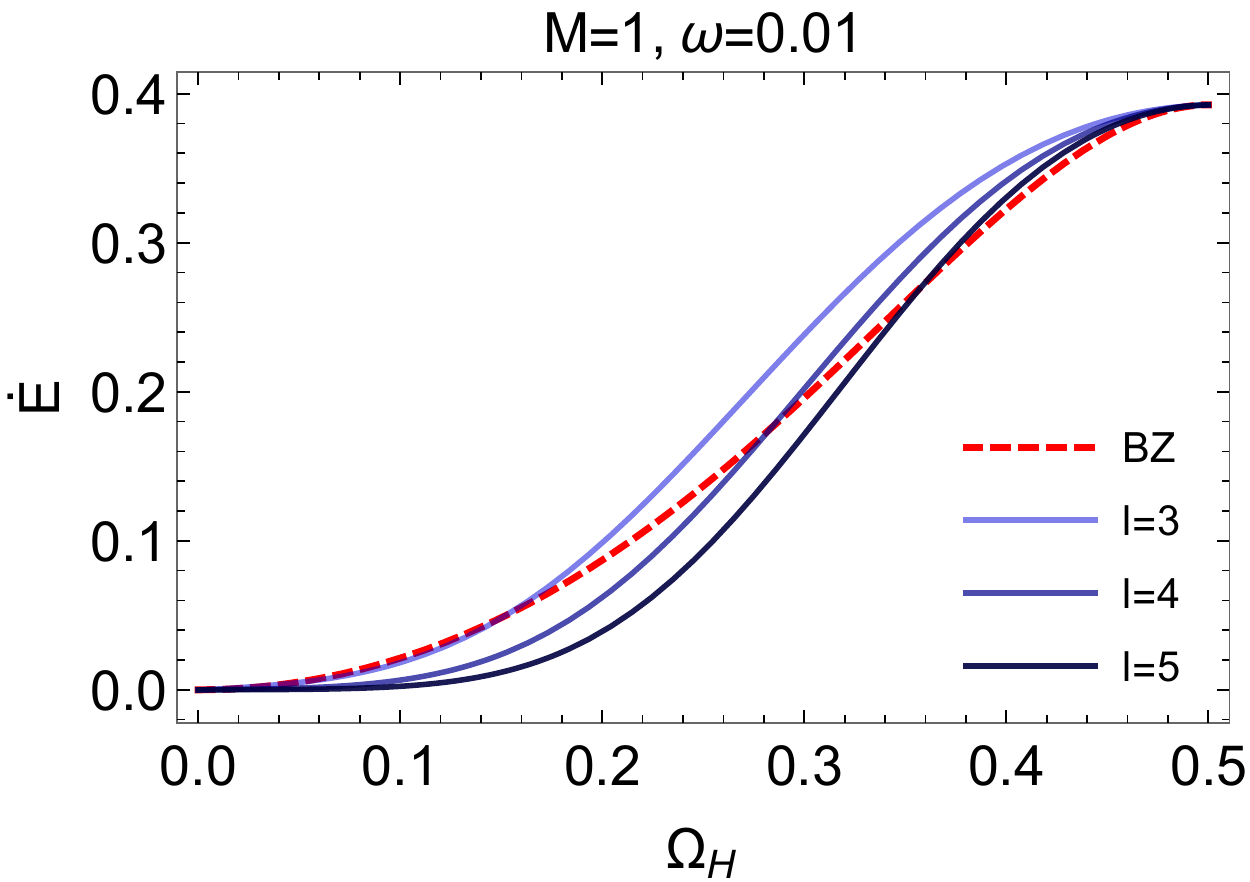}
  \includegraphics[width=0.95\linewidth]{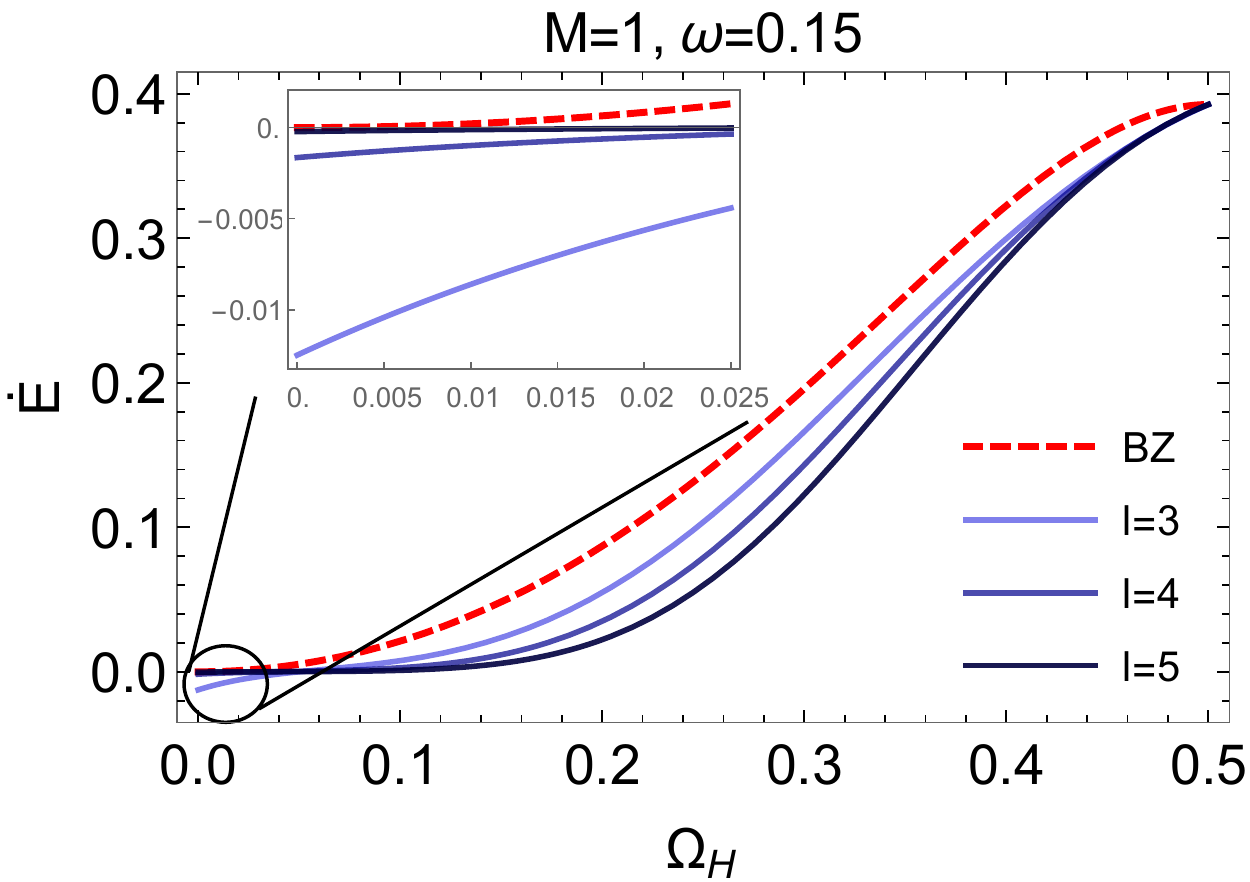}
\caption{\label{fig:fluxAgainstSpin} \footnotesize Energy extraction rates plotted against horizon angular velocity for two scalar rotation frequencies, $\omega M = 0.01$ and $0.15$. The dashed line is the $\mathcal{O}(\Omega_{H}^{8})$ prediction of the BZ mechanism presented in~\cite{Pan:2015iaa}. The solid lines are the results from (\ref{Edot}) plotted for various harmonics $l=m$. All curves are normalized to reach the same maximum value as the BZ curve.}
\label{fig:energyAgainstSpin}
\end{figure}

As shown in Fig.~\ref{fig:fluxAgainstSpin}, the energy extraction rate increases monotonically with the horizon angular velocity $\Omega_{H}$ because the black hole is effectively a stronger power source. In the slow rotation limit the energy extraction rate is negative because the superradiant condition is not met. This behavior does not occur in the BZ mechanism where instead $\dot{E}\to 0$ in the slow rotation limit. Since we've only considered small $\omega M$, energy is extracted for most of the spin range and the qualitative similarities between the scalar solution and the BZ mechanism are more obvious.  Both our curves and the plotted BZ curve  increase slowly for small spins, increase fairly linearly for intermediate spins, and saturate to a constant value near maximal spins.  This saturation behaviour has been seen in GRMHD simulations of the BZ process, but until~\cite{Pan:2015iaa} it had not been predicted by theoretical calculation.  Our calculation is valid for the entire range of spins and also displays this saturation behaviour, which is suggestive that  this qualitative behaviour of the power curve is universal for stationary BH energy extraction. 

To analyze the similarity of the power curves in more detail, we develop expansions of (\ref{Edot}) in the slow-rotation and near-extremal limits, taking $M=1,\omega=0.01,l=m=3$ as an example. The slow rotation expansion, with the amplitude normalized as detailed above, takes the form
\begin{align} \label{eq:slowrotexp}
\dot{E} &=  \, 0.08\Omega_{H}-0.15\Omega_{H}^2+11.75\Omega_{H}^3  -4.28\Omega_{H}^4+64.48 \Omega_{H} ^5 \nonumber \\
& \;\;+84.52\Omega_{H}^6-3013.13\Omega_{H}^7 -734.71\Omega_{H}^8+\mathcal{O}\left(\Omega_{H}^9\right).
\end{align}
This can be compared with the expansion for the BZ radial power \cite{Pan:2015iaa},
\begin{align} \label{BZexp}
\dot{E}_{\rm BZ} &= 2.09 \Omega_{H}^2+2.89 \Omega_{H}^4-23.23 \Omega_{H}^6 \nonumber\\
 & \qquad +13.11 \Omega_{H}^8 +\mathcal{O}\left(\Omega_{H}^{10} \right).
\end{align}
The curves (\ref{eq:slowrotexp}) and (\ref{BZexp}) agree well for $\Omega_{H}\lesssim 0.2$. The primary distinction is the presence of only even powers of $\Omega_H$ in the latter case, as a consequence of the structure of the electromagnetic energy momentum tensor. 

We also show the corresponding expansion in the near-extremal limit, in powers of $\zeta=\Omega_{H}-\Omega_H^{\rm ext}$, where the extremal limit in 
these units corresponds to $\Omega_H^{\rm ext}=\frac{1}{2}$,
\begin{align}
\dot{E} & = 0.3925+0.0369 \zeta-3.092 \zeta^2
 +5.83 \zeta^3 +4.71 \zeta^4 \nonumber\\
  & \;\; -40.02 \zeta^5+70.28 \zeta^6+10.27 \zeta^7-313.99 \zeta^8+\mathcal{O}(\zeta^9 ),
\end{align}
This clearly exhibits extremal saturation, with the approach to the limit being almost quadratic in $\zeta$ due to suppression of the linear term. The scaling is similar to the BZ curve for $\Omega_{H}\gtrsim 0.2$. 

\section{\label{sec:Conclusions}Concluding Remarks}

We have studied stationary scalar cloud configurations around Kerr black holes as tractable proxies for the force-free magnetosphere within the BZ mechanism. While local perturbations of the scalar cloud tend to move the existence curve for the solution around in parameter space, we found that introducing a semi-permeable membrane outside the ergosphere could, with a tuned transmission coefficient, lead to a stationary radial energy flux. In detail, we solved the scalar field equation analytically in the long-wavelength limit, $\omega M\ll 1$, which allowed a calculation of the tuning condition required to cancel the imaginary parts of the quasinormal modes. This tuning is similar to the threshold existence condition for scalar clouds, as it lies in between the decaying regime of superradiant scattering and the exponential instability induced in the presence of a perfectly reflecting mirror. The rate of energy extraction was calculated for the entire range of BH spins and compared to the most recent results for the BZ split monopole solution. The curves $\dot{E}(\Omega_{H})$ obtained from the scalar field proxy were qualitatively quite similar to those for the BZ process (e.g. as recently obtained in~\cite{Pan:2015iaa}), especially in the physically interesting near-extremal regime. 

Throughout, these computations were performed in the probe limit, ignoring backreaction on the metric. This is a physically motivated assumption for the BZ process, where the energy density outside the black hole is generally assumed to be negligible compared to the impact of the BH itself. In the present case, one can also ask about the backreaction of the semi-permeable membrane introduced to generate the stationary energy flux. The boundary conditions ensure that this surface carries finite scalar charge density, and thus energy density. Its backreaction could be determined by imposing the appropriate Israel junction conditions, and it would be interesting to explore this in more detail. In the full theory, it would be interesting to see if stationarity could be maintained with this membrane sitting at the (spherically or axially symmetric) analogue of `floating' orbits \cite{PhysRevLett.28.994}. It would also be interesting to study the energy extraction rate further in the near-extremal limit, where there was a close resemblance to the curve for the BZ process, as this may provide physical intuition into the physics of the BZ mechanism for astrophysically relevant rapidly-rotating black holes. Finally, it would also be interesting to study the analogue of this system with anti-de Sitter boundary conditions, where the holographic correspondence may allow further insight into classical energy extraction from rotating black holes \cite{Wang:2014vza}.

\begin{acknowledgments}
This work was supported 
in part by NSERC, Canada. 
\end{acknowledgments}

\bibliography{energyextraction.bib}

\begin{thebibliography}{26}%
\makeatletter
\providecommand \@ifxundefined [1]{%
 \@ifx{#1\undefined}
}%
\providecommand \@ifnum [1]{%
 \ifnum #1\expandafter \@firstoftwo
 \else \expandafter \@secondoftwo
 \fi
}%
\providecommand \@ifx [1]{%
 \ifx #1\expandafter \@firstoftwo
 \else \expandafter \@secondoftwo
 \fi
}%
\providecommand \natexlab [1]{#1}%
\providecommand \enquote  [1]{``#1''}%
\providecommand \bibnamefont  [1]{#1}%
\providecommand \bibfnamefont [1]{#1}%
\providecommand \citenamefont [1]{#1}%
\providecommand \href@noop [0]{\@secondoftwo}%
\providecommand \href [0]{\begingroup \@sanitize@url \@href}%
\providecommand \@href[1]{\@@startlink{#1}\@@href}%
\providecommand \@@href[1]{\endgroup#1\@@endlink}%
\providecommand \@sanitize@url [0]{\catcode `\\12\catcode `\$12\catcode
  `\&12\catcode `\#12\catcode `\^12\catcode `\_12\catcode `\%12\relax}%
\providecommand \@@startlink[1]{}%
\providecommand \@@endlink[0]{}%
\providecommand \url  [0]{\begingroup\@sanitize@url \@url }%
\providecommand \@url [1]{\endgroup\@href {#1}{\urlprefix }}%
\providecommand \urlprefix  [0]{URL }%
\providecommand \Eprint [0]{\href }%
\providecommand \doibase [0]{http://dx.doi.org/}%
\providecommand \selectlanguage [0]{\@gobble}%
\providecommand \bibinfo  [0]{\@secondoftwo}%
\providecommand \bibfield  [0]{\@secondoftwo}%
\providecommand \translation [1]{[#1]}%
\providecommand \BibitemOpen [0]{}%
\providecommand \bibitemStop [0]{}%
\providecommand \bibitemNoStop [0]{.\EOS\space}%
\providecommand \EOS [0]{\spacefactor3000\relax}%
\providecommand \BibitemShut  [1]{\csname bibitem#1\endcsname}%
\let\auto@bib@innerbib\@empty
\bibitem [{\citenamefont {Penrose}\ and\ \citenamefont
  {Floyd}(1971)}]{Penrose:1971uk}%
  \BibitemOpen
  \bibfield  {author} {\bibinfo {author} {\bibfnamefont {R.}~\bibnamefont
  {Penrose}}\ and\ \bibinfo {author} {\bibfnamefont {R.~M.}\ \bibnamefont
  {Floyd}},\ }\href@noop {} {\bibfield  {journal} {\bibinfo  {journal}
  {Nature}\ }\textbf {\bibinfo {volume} {229}},\ \bibinfo {pages} {177}
  (\bibinfo {year} {1971})}\BibitemShut {NoStop}%
\bibitem [{\citenamefont {Bekenstein}(1973)}]{Bekenstein:1973mi}%
  \BibitemOpen
  \bibfield  {author} {\bibinfo {author} {\bibfnamefont {J.~D.}\ \bibnamefont
  {Bekenstein}},\ }\href {\doibase 10.1103/PhysRevD.7.949} {\bibfield
  {journal} {\bibinfo  {journal} {Phys. Rev.}\ }\textbf {\bibinfo {volume}
  {D7}},\ \bibinfo {pages} {949} (\bibinfo {year} {1973})}\BibitemShut
  {NoStop}%
\bibitem [{\citenamefont {Misner}(1972)}]{PhysRevLett.28.994}%
  \BibitemOpen
  \bibfield  {author} {\bibinfo {author} {\bibfnamefont {C.~W.}\ \bibnamefont
  {Misner}},\ }\href {\doibase 10.1103/PhysRevLett.28.994} {\bibfield
  {journal} {\bibinfo  {journal} {Phys. Rev. Lett.}\ }\textbf {\bibinfo
  {volume} {28}},\ \bibinfo {pages} {994} (\bibinfo {year} {1972})}\BibitemShut
  {NoStop}%
\bibitem [{\citenamefont {Zel'dovich}(1971)}]{JETP.Let.14.180}%
  \BibitemOpen
  \bibfield  {author} {\bibinfo {author} {\bibfnamefont {Y.~B.}\ \bibnamefont
  {Zel'dovich}},\ }\href@noop {} {\bibfield  {journal} {\bibinfo  {journal}
  {Zh. Eksp. Teor. Fiz. Pis'ma}\ }\textbf {\bibinfo {volume} {14}},\ \bibinfo
  {pages} {270} (\bibinfo {year} {1971})}\BibitemShut {NoStop}%
\bibitem [{\citenamefont {Hawking}(1972)}]{Hawking:1971vc}%
  \BibitemOpen
  \bibfield  {author} {\bibinfo {author} {\bibfnamefont {S.~W.}\ \bibnamefont
  {Hawking}},\ }\href {\doibase 10.1007/BF01877517} {\bibfield  {journal}
  {\bibinfo  {journal} {Commun. Math. Phys.}\ }\textbf {\bibinfo {volume}
  {25}},\ \bibinfo {pages} {152} (\bibinfo {year} {1972})}\BibitemShut
  {NoStop}%
\bibitem [{\citenamefont {Brito}\ \emph {et~al.}(2015)\citenamefont {Brito},
  \citenamefont {Cardoso},\ and\ \citenamefont {Pani}}]{Brito:2015oca}%
  \BibitemOpen
  \bibfield  {author} {\bibinfo {author} {\bibfnamefont {R.}~\bibnamefont
  {Brito}}, \bibinfo {author} {\bibfnamefont {V.}~\bibnamefont {Cardoso}}, \
  and\ \bibinfo {author} {\bibfnamefont {P.}~\bibnamefont {Pani}},\ }\href@noop
  {} {\  (\bibinfo {year} {2015})},\ \Eprint {http://arxiv.org/abs/1501.06570}
  {arXiv:1501.06570 [gr-qc]} \BibitemShut {NoStop}%
\bibitem [{\citenamefont {Blandford}\ and\ \citenamefont
  {Znajek}(1977)}]{Blandford:1977ds}%
  \BibitemOpen
  \bibfield  {author} {\bibinfo {author} {\bibfnamefont {R.~D.}\ \bibnamefont
  {Blandford}}\ and\ \bibinfo {author} {\bibfnamefont {R.~L.}\ \bibnamefont
  {Znajek}},\ }\href@noop {} {\bibfield  {journal} {\bibinfo  {journal} {Mon.
  Not. Roy. Astron. Soc.}\ }\textbf {\bibinfo {volume} {179}},\ \bibinfo
  {pages} {433} (\bibinfo {year} {1977})}\BibitemShut {NoStop}%
\bibitem [{\citenamefont {Tanabe}\ and\ \citenamefont
  {Nagataki}(2008)}]{Tanabe:2008wm}%
  \BibitemOpen
  \bibfield  {author} {\bibinfo {author} {\bibfnamefont {K.}~\bibnamefont
  {Tanabe}}\ and\ \bibinfo {author} {\bibfnamefont {S.}~\bibnamefont
  {Nagataki}},\ }\href {\doibase 10.1103/PhysRevD.78.024004} {\bibfield
  {journal} {\bibinfo  {journal} {Phys. Rev.}\ }\textbf {\bibinfo {volume}
  {D78}},\ \bibinfo {pages} {024004} (\bibinfo {year} {2008})},\ \Eprint
  {http://arxiv.org/abs/0802.0908} {arXiv:0802.0908 [astro-ph]} \BibitemShut
  {NoStop}%
\bibitem [{\citenamefont {Komissarov}(2004)}]{Komissarov:2004qu}%
  \BibitemOpen
  \bibfield  {author} {\bibinfo {author} {\bibfnamefont {S.~S.}\ \bibnamefont
  {Komissarov}},\ }\href {\doibase 10.1111/j.1365-2966.2004.07738.x} {\bibfield
   {journal} {\bibinfo  {journal} {Mon. Not. Roy. Astron. Soc.}\ }\textbf
  {\bibinfo {volume} {350}},\ \bibinfo {pages} {1431} (\bibinfo {year}
  {2004})},\ \Eprint {http://arxiv.org/abs/astro-ph/0402430}
  {arXiv:astro-ph/0402430 [astro-ph]} \BibitemShut {NoStop}%
\bibitem [{\citenamefont {McKinney}\ and\ \citenamefont
  {Gammie}(2004)}]{McKinney:2004ka}%
  \BibitemOpen
  \bibfield  {author} {\bibinfo {author} {\bibfnamefont {J.~C.}\ \bibnamefont
  {McKinney}}\ and\ \bibinfo {author} {\bibfnamefont {C.~F.}\ \bibnamefont
  {Gammie}},\ }\href {\doibase 10.1086/422244} {\bibfield  {journal} {\bibinfo
  {journal} {Astrophys. J.}\ }\textbf {\bibinfo {volume} {611}},\ \bibinfo
  {pages} {977} (\bibinfo {year} {2004})},\ \Eprint
  {http://arxiv.org/abs/astro-ph/0404512} {arXiv:astro-ph/0404512 [astro-ph]}
  \BibitemShut {NoStop}%
\bibitem [{\citenamefont {Tchekhovskoy}\ \emph {et~al.}(2010)\citenamefont
  {Tchekhovskoy}, \citenamefont {Narayan},\ and\ \citenamefont
  {McKinney}}]{Tchekhovskoy:2009ba}%
  \BibitemOpen
  \bibfield  {author} {\bibinfo {author} {\bibfnamefont {A.}~\bibnamefont
  {Tchekhovskoy}}, \bibinfo {author} {\bibfnamefont {R.}~\bibnamefont
  {Narayan}}, \ and\ \bibinfo {author} {\bibfnamefont {J.~C.}\ \bibnamefont
  {McKinney}},\ }\href {\doibase 10.1088/0004-637X/711/1/50} {\bibfield
  {journal} {\bibinfo  {journal} {Astrophys. J.}\ }\textbf {\bibinfo {volume}
  {711}},\ \bibinfo {pages} {50} (\bibinfo {year} {2010})},\ \Eprint
  {http://arxiv.org/abs/0911.2228} {arXiv:0911.2228 [astro-ph.HE]} \BibitemShut
  {NoStop}%
\bibitem [{\citenamefont {Pan}\ and\ \citenamefont {Yu}(2015)}]{Pan:2015iaa}%
  \BibitemOpen
  \bibfield  {author} {\bibinfo {author} {\bibfnamefont {Z.}~\bibnamefont
  {Pan}}\ and\ \bibinfo {author} {\bibfnamefont {C.}~\bibnamefont {Yu}},\
  }\href@noop {} {\  (\bibinfo {year} {2015})},\ \Eprint
  {http://arxiv.org/abs/1504.04864} {arXiv:1504.04864 [astro-ph.HE]}
  \BibitemShut {NoStop}%
\bibitem [{\citenamefont {Brennan}\ \emph {et~al.}(2013)\citenamefont
  {Brennan}, \citenamefont {Gralla},\ and\ \citenamefont {Jacobson}}]{BGJ}%
  \BibitemOpen
  \bibfield  {author} {\bibinfo {author} {\bibfnamefont {T.~D.}\ \bibnamefont
  {Brennan}}, \bibinfo {author} {\bibfnamefont {S.~E.}\ \bibnamefont {Gralla}},
  \ and\ \bibinfo {author} {\bibfnamefont {T.}~\bibnamefont {Jacobson}},\
  }\href {\doibase 10.1088/0264-9381/30/19/195012} {\bibfield  {journal}
  {\bibinfo  {journal} {Class. Quant. Grav.}\ }\textbf {\bibinfo {volume}
  {30}},\ \bibinfo {pages} {195012} (\bibinfo {year} {2013})},\ \Eprint
  {http://arxiv.org/abs/1305.6890} {arXiv:1305.6890 [gr-qc]} \BibitemShut
  {NoStop}%
\bibitem [{\citenamefont {Menon}\ and\ \citenamefont {Dermer}(2007)}]{MD}%
  \BibitemOpen
  \bibfield  {author} {\bibinfo {author} {\bibfnamefont {G.}~\bibnamefont
  {Menon}}\ and\ \bibinfo {author} {\bibfnamefont {C.~D.}\ \bibnamefont
  {Dermer}},\ }\href {\doibase 10.1007/s10714-007-0418-2} {\bibfield  {journal}
  {\bibinfo  {journal} {Gen. Rel. Grav.}\ }\textbf {\bibinfo {volume} {39}},\
  \bibinfo {pages} {785} (\bibinfo {year} {2007})},\ \Eprint
  {http://arxiv.org/abs/astro-ph/0511661} {arXiv:astro-ph/0511661 [astro-ph]}
  \BibitemShut {NoStop}%
\bibitem [{\citenamefont {Hod}(2012)}]{Hod:2012px}%
  \BibitemOpen
  \bibfield  {author} {\bibinfo {author} {\bibfnamefont {S.}~\bibnamefont
  {Hod}},\ }\href {\doibase 10.1103/PhysRevD.86.129902,
  10.1103/PhysRevD.86.104026} {\bibfield  {journal} {\bibinfo  {journal} {Phys.
  Rev.}\ }\textbf {\bibinfo {volume} {D86}},\ \bibinfo {pages} {104026}
  (\bibinfo {year} {2012})},\ \bibinfo {note} {[Erratum: Phys.
  Rev.D86,129902(2012)]},\ \Eprint {http://arxiv.org/abs/1211.3202}
  {arXiv:1211.3202 [gr-qc]} \BibitemShut {NoStop}%
\bibitem [{\citenamefont {Herdeiro}\ and\ \citenamefont
  {Radu}(2014)}]{Herdeiro:2014goa}%
  \BibitemOpen
  \bibfield  {author} {\bibinfo {author} {\bibfnamefont {C.~A.~R.}\
  \bibnamefont {Herdeiro}}\ and\ \bibinfo {author} {\bibfnamefont
  {E.}~\bibnamefont {Radu}},\ }\href {\doibase 10.1103/PhysRevLett.112.221101}
  {\bibfield  {journal} {\bibinfo  {journal} {Phys. Rev. Lett.}\ }\textbf
  {\bibinfo {volume} {112}},\ \bibinfo {pages} {221101} (\bibinfo {year}
  {2014})},\ \Eprint {http://arxiv.org/abs/1403.2757} {arXiv:1403.2757 [gr-qc]}
  \BibitemShut {NoStop}%
\bibitem [{\citenamefont {Hod}(2014)}]{HodEM}%
  \BibitemOpen
  \bibfield  {author} {\bibinfo {author} {\bibfnamefont {S.}~\bibnamefont
  {Hod}},\ }\href {\doibase 10.1103/PhysRevD.90.024051} {\bibfield  {journal}
  {\bibinfo  {journal} {Phys. Rev.}\ }\textbf {\bibinfo {volume} {D90}},\
  \bibinfo {pages} {024051} (\bibinfo {year} {2014})},\ \Eprint
  {http://arxiv.org/abs/1406.1179} {arXiv:1406.1179 [gr-qc]} \BibitemShut
  {NoStop}%
\bibitem [{\citenamefont {Benone}\ \emph {et~al.}(2014)\citenamefont {Benone},
  \citenamefont {Crispino}, \citenamefont {Herdeiro},\ and\ \citenamefont
  {Radu}}]{BenoneEM}%
  \BibitemOpen
  \bibfield  {author} {\bibinfo {author} {\bibfnamefont {C.~L.}\ \bibnamefont
  {Benone}}, \bibinfo {author} {\bibfnamefont {L.~C.~B.}\ \bibnamefont
  {Crispino}}, \bibinfo {author} {\bibfnamefont {C.}~\bibnamefont {Herdeiro}},
  \ and\ \bibinfo {author} {\bibfnamefont {E.}~\bibnamefont {Radu}},\ }\href
  {\doibase 10.1103/PhysRevD.90.104024} {\bibfield  {journal} {\bibinfo
  {journal} {Phys. Rev.}\ }\textbf {\bibinfo {volume} {D90}},\ \bibinfo {pages}
  {104024} (\bibinfo {year} {2014})},\ \Eprint {http://arxiv.org/abs/1409.1593}
  {arXiv:1409.1593 [gr-qc]} \BibitemShut {NoStop}%
\bibitem [{\citenamefont {Press}\ and\ \citenamefont
  {Teukolsky}(1972)}]{Press:1972zz}%
  \BibitemOpen
  \bibfield  {author} {\bibinfo {author} {\bibfnamefont {W.~H.}\ \bibnamefont
  {Press}}\ and\ \bibinfo {author} {\bibfnamefont {S.~A.}\ \bibnamefont
  {Teukolsky}},\ }\href {\doibase 10.1038/238211a0} {\bibfield  {journal}
  {\bibinfo  {journal} {Nature}\ }\textbf {\bibinfo {volume} {238}},\ \bibinfo
  {pages} {211} (\bibinfo {year} {1972})}\BibitemShut {NoStop}%
\bibitem [{\citenamefont {Cardoso}\ \emph {et~al.}(2004)\citenamefont
  {Cardoso}, \citenamefont {Dias}, \citenamefont {Lemos},\ and\ \citenamefont
  {Yoshida}}]{Cardoso:2004nk}%
  \BibitemOpen
  \bibfield  {author} {\bibinfo {author} {\bibfnamefont {V.}~\bibnamefont
  {Cardoso}}, \bibinfo {author} {\bibfnamefont {O.~J.}\ \bibnamefont {Dias}},
  \bibinfo {author} {\bibfnamefont {J.~P.~S.}\ \bibnamefont {Lemos}}, \ and\
  \bibinfo {author} {\bibfnamefont {S.}~\bibnamefont {Yoshida}},\ }\href
  {\doibase 10.1103/PhysRevD.70.049903, 10.1103/PhysRevD.70.044039} {\bibfield
  {journal} {\bibinfo  {journal} {Phys. Rev.}\ }\textbf {\bibinfo {volume}
  {D70}},\ \bibinfo {pages} {044039} (\bibinfo {year} {2004})},\ \bibinfo
  {note} {[Erratum: Phys. Rev.D70,049903(2004)]},\ \Eprint
  {http://arxiv.org/abs/hep-th/0404096} {arXiv:hep-th/0404096 [hep-th]}
  \BibitemShut {NoStop}%
\bibitem [{\citenamefont {Cardoso}\ and\ \citenamefont
  {Dias}(2004)}]{PhysRevD.70.084011}%
  \BibitemOpen
  \bibfield  {author} {\bibinfo {author} {\bibfnamefont {V.}~\bibnamefont
  {Cardoso}}\ and\ \bibinfo {author} {\bibfnamefont {O.~J.~C.}\ \bibnamefont
  {Dias}},\ }\href {\doibase 10.1103/PhysRevD.70.084011} {\bibfield  {journal}
  {\bibinfo  {journal} {Phys. Rev. D}\ }\textbf {\bibinfo {volume} {70}},\
  \bibinfo {pages} {084011} (\bibinfo {year} {2004})}\BibitemShut {NoStop}%
\bibitem [{\citenamefont {Brito}\ \emph {et~al.}(2014)\citenamefont {Brito},
  \citenamefont {Cardoso},\ and\ \citenamefont {Pani}}]{PhysRevD.89.104045}%
  \BibitemOpen
  \bibfield  {author} {\bibinfo {author} {\bibfnamefont {R.}~\bibnamefont
  {Brito}}, \bibinfo {author} {\bibfnamefont {V.}~\bibnamefont {Cardoso}}, \
  and\ \bibinfo {author} {\bibfnamefont {P.}~\bibnamefont {Pani}},\ }\href
  {\doibase 10.1103/PhysRevD.89.104045} {\bibfield  {journal} {\bibinfo
  {journal} {Phys. Rev. D}\ }\textbf {\bibinfo {volume} {89}},\ \bibinfo
  {pages} {104045} (\bibinfo {year} {2014})}\BibitemShut {NoStop}%
\bibitem [{\citenamefont {Damour}\ \emph {et~al.}(1976)\citenamefont {Damour},
  \citenamefont {Deruelle},\ and\ \citenamefont {Ruffini}}]{damour}%
  \BibitemOpen
  \bibfield  {author} {\bibinfo {author} {\bibfnamefont {T.}~\bibnamefont
  {Damour}}, \bibinfo {author} {\bibfnamefont {N.}~\bibnamefont {Deruelle}}, \
  and\ \bibinfo {author} {\bibfnamefont {R.}~\bibnamefont {Ruffini}},\ }\href
  {\doibase 10.1007/BF02725534} {\bibfield  {journal} {\bibinfo  {journal}
  {Lettere al Nuovo Cimento (1971-1985)}\ }\textbf {\bibinfo {volume} {15}},\
  \bibinfo {pages} {257} (\bibinfo {year} {1976})}\BibitemShut {NoStop}%
\bibitem [{\citenamefont {Hod}\ and\ \citenamefont {Hod}(2010)}]{Hod:2009cp}%
  \BibitemOpen
  \bibfield  {author} {\bibinfo {author} {\bibfnamefont {S.}~\bibnamefont
  {Hod}}\ and\ \bibinfo {author} {\bibfnamefont {O.}~\bibnamefont {Hod}},\
  }\href {\doibase 10.1103/PhysRevD.81.061502} {\bibfield  {journal} {\bibinfo
  {journal} {Phys. Rev.}\ }\textbf {\bibinfo {volume} {D81}},\ \bibinfo {pages}
  {061502} (\bibinfo {year} {2010})},\ \Eprint {http://arxiv.org/abs/0910.0734}
  {arXiv:0910.0734 [gr-qc]} \BibitemShut {NoStop}%
\bibitem [{\citenamefont {Abramowitz}\ and\ \citenamefont
  {Stegun}(1970)}]{abramowitz}%
  \BibitemOpen
  \bibfield  {author} {\bibinfo {author} {\bibfnamefont {M.}~\bibnamefont
  {Abramowitz}}\ and\ \bibinfo {author} {\bibfnamefont {A.}~\bibnamefont
  {Stegun}},\ }\href@noop {} {\emph {\bibinfo {title} {Handbook of mathematical
  functions}}}\ (\bibinfo  {publisher} {Dover Publications, New York},\
  \bibinfo {year} {1970})\BibitemShut {NoStop}%
\bibitem [{\citenamefont {Wang}\ and\ \citenamefont
  {Ritz}(2014)}]{Wang:2014vza}%
  \BibitemOpen
  \bibfield  {author} {\bibinfo {author} {\bibfnamefont {X.}~\bibnamefont
  {Wang}}\ and\ \bibinfo {author} {\bibfnamefont {A.}~\bibnamefont {Ritz}},\
  }\href {\doibase 10.1103/PhysRevD.89.106011} {\bibfield  {journal} {\bibinfo
  {journal} {Phys. Rev.}\ }\textbf {\bibinfo {volume} {D89}},\ \bibinfo {pages}
  {106011} (\bibinfo {year} {2014})},\ \Eprint {http://arxiv.org/abs/1402.1452}
  {arXiv:1402.1452 [hep-th]} \BibitemShut {NoStop}%
\end{thebibliography}%

\end{document}